\documentclass[aps,prb,twocolumn,showpacs,superscriptaddress,floatfix]{revtex4}
\usepackage{amsmath,amssymb}
\usepackage{graphicx}

\newcommand{\bfq}{{\mathbf{q}}}
\newcommand{\bfr}{{\mathbf{r}}}
\newcommand{\bfx}{{\mathbf{x}}}
\newcommand{\bfy}{{\mathbf{y}}}
\newcommand{\bfz}{{\mathbf{z}}}

\newcommand{\bfS}{{\mathbf{S}}}
\newcommand{\bfsigma}{{\boldsymbol{\sigma}}}
\newcommand{\bftau}{{\boldsymbol{\tau}}}

\newcommand{\varD}{{\mathcal{D}}}
\newcommand{\varH}{{\mathcal{H}}}
\newcommand{\varK}{{\mathcal{K}}}
\newcommand{\varO}{{\mathcal{O}}}
\newcommand{\varZ}{{\mathcal{Z}}}

\newcommand{\ux}{{\hat\bfx}}
\newcommand{\uy}{{\hat\bfy}}
\newcommand{\uz}{{\hat\bfz}}

\newcommand{\pde}[2]{\frac{\partial{#1}}{\partial{#2}}}

\newcommand{\half}{\frac{1}{2}}
\newcommand{\rcp}[1]{\frac{1}{#1}}

\newcommand{\avg}[1]{\left\langle#1\right\rangle}
\newcommand{\nn}[1]{\left\langle#1\right\rangle}

\newcommand{\eqnref}[1]{Eq.~(\ref{#1})}
\newcommand{\eqnsref}[1]{Eqs.~(\ref{#1})}

\newcommand{\figref}[1]{Fig.~\ref{#1}}
\newcommand{\Figref}[1]{Figure~\ref{#1}}
\newcommand{\secref}[1]{Sec.~\ref{#1}}
\newcommand{\secsref}[1]{Secs.~\ref{#1}}
\newcommand{\Secref}[1]{Section~\ref{#1}}

\newcommand{\XY}{\textit{XY} }

\begin{document}

\title{Phase Transitions in Models for Coupled Charge-Density Waves}

\author{Minchul Lee}
\affiliation{Department of Physics, Korea University, Seoul 136-701, Korea}

\author{Eun-Ah Kim}
\affiliation{Department of Physics, University of Illinois, Champaign, IL 61801, U.S.A.}

\author{Jong Soo Lim}
\affiliation{Department of Physics, Seoul National University, Seoul 151-747, Korea}

\author{M. Y. Choi}
\affiliation{Department of Physics, Seoul National University, Seoul 151-747, Korea}
\affiliation{Korea Institute for Advanced Study, Seoul 130-722, Korea}

\begin{abstract}
Various phase transitions in models for coupled charge-density waves are investigated
by means of the $\epsilon$-expansion, mean-field theory, and Monte Carlo simulations.
At zero temperature the effective action for the system with appropriate commensurability
effects is mapped onto the three- or four-dimensional \XY model, depending on
spatiotemporal fluctuations, under the corresponding symmetry-breaking fields.
It is revealed that the three- and four-dimensional systems display a single transition between
the clock order (with broken Z$_M$ symmetry) and disorder.
The nature of the phase transition depends crucially on the commensurability factor $M$:
For $M \ge 4$, in particular, the transition belongs to the same university class as the \XY model.
On the other hand, in the presence of misfit causing frustration in the charge-density wave,
the inter-chain coupling is observed to favor either the commensurate state or the
incommensurate state depending on the initial configuration; this gives rise to hysteresis around
the commensurate-incommensurate transition.
Boundaries separating such phases with different symmetries are obtained in the parameter
space consisting of the temperature, symmetry-breaking field, fluctuation strength,
inter-chain coupling, and misfit.
\end{abstract}
\bigskip
\pacs{64.60.Cn, 71.45.Lr, 64.70.Rh}

\maketitle

\section{Introduction}

A large number of organic and inorganic solids have crystalline structures in which
fundamental structural units form linear chains.\cite{Kagoshima87}  In these materials,
largely different overlaps of the electronic wave functions in various crystallographic
directions lead to strongly anisotropic, so-called quasi-one-dimensional (quasi-1D)
electron bands.  Among the exotic physical phenomena in quasi-1D materials, charge density waves
(CDWs) are of much continual interest.  Research into this topic has been stimulated
by recent advances in experimental techniques, which now allow direct observation of
CDWs and measurement of various static and dynamic properties of CDW systems.\cite{Experiment}
In general the material of quasi-1D structure is expected to exist in the form of
a bundle of chains rather than of a single chain.  In such a bundle of chains,
inter-chain tunneling of electrons leads to coupling of the fluctuations on
neighboring chains, which may affect the behavior of the system in a crucial way.
For example, the system of two coupled incommensurate chains in the weak-coupling limit
has been reported to exhibit a complicated commensurate-incommensurate (CI) transition,
reminiscent of devil's staircase.\cite{Coutinho81}
Obviously, the opening of the gap at the Fermi surface implies that each separate chain is an insulator
at low temperatures.  However, with the coupling between chains taken into account, expected are
various transitions between insulating and metallic phases, the extensive investigation of which is still
lacking in spite of the ubiquity of the CDW.

In this paper, we investigate nature of the phase transitions in the coupled CDW system.
In the absence of misfit the effective action for the (commensurate) system at zero temperature
with suitable commensurability effects is mapped onto the \XY model under the corresponding
symmetry-breaking field.  The effective dimension of the system is given by four if both spatial
and temporal fluctuations are significant; otherwise it is three.
The phase transition in the resulting three-dimensional (3D) \XY model under the appropriate
symmetry-breaking field is examined via the $\epsilon$-expansion and the Monte Carlo method.
It is found that there emerges in the system the clock ordered phase (with the Z$_M$ symmetry broken)
via a second-order transition, the nature of which depends on the commensurability factor $M$:
In particular the critical behavior for $M\ge4$ appears to be the same as that of the 3D \XY model.
For the four-dimensional (4D) \XY model, on the other hand, mean-field theory is expected to be accurate
and reveals a single transition, which is of the first order for $M=3$ and of the second order otherwise.
We then examine the effects of misfit, which not only change the nature of the order-disorder transition
but also brings about a CI transition.  In the presence of the misfit, correlations due to the inter-chain
coupling are observed to favor, depending on the initial configuration, either the commensurate state or an
incommensurate state, which gives rise to hysteresis behavior:
While in the cooling process the incommensurate CDWs persist near zero temperature, in
the heating process large portions of the system remain in the commensurate state
at rather high temperatures.

This paper is organized as follows: In \secref{sec:ea}, the effective action at zero temperature is derived
for the coupled CDW system and mapped onto the appropriate \XY model according to whether
spatial and/or temporal fluctuations are taken into consideration.  \Secref{sec:3dxy} is devoted to
the investigation of the phase transition in the 3D \XY model under the symmetry-breaking field,
which introduces the Z$_M$ symmetry to the system.  Here two independent approaches are
employed: the $\epsilon$-expansion in \secref{sec:eexp} and the Monte Carlo method in \secref{sec:3dxymc}.
\Secref{sec:4dxy} presents the mean-field analysis of the 4D \XY model, demonstrating the first- and
second-order nature of the transition under the appropriate symmetry-breaking field.
In \secref{sec:cic} the effects of the inter-chain coupling in the presence of misfit are investigated
via Monte Carlo simulations, which discloses properties of the CI transition in the system.
Combining the results for the CI transition with those for the 3D and 4D \XY models
obtained in \secref{sec:3dxy} and \ref{sec:4dxy}, we construct in \secref{sec:pd} schematic
phase diagrams for general coupled CDW systems in the 3D space consisting of
the temperature, inter-chain coupling, and misfit.  Finally, \secref{sec:c} gives a brief summary.

%
\section{Effective Action\label{sec:ea}}

We consider a system of coupled near-commensurate CDW chains along the $z$ direction,
each of which is characterized by the commensurability factor $M$ and the position-dependent
misfit $\delta_\bfr$ with $\bfr=(x,y,z)$.  On the $x$-$y$ plane, the chains are assumed for simplicity
to constitute a square array of lattice constant $a\, (\equiv 1)$.
Disregarding amplitude fluctuations of the complex CDW order parameter and
considering spatial and temporal fluctuations of the phase only,\cite{Gruner94}
we write the Hamiltonian in terms of the phase $\phi_\bfr$ of the order parameter at
position $\bfr$ and the momentum $p_\bfr = C\partial\phi_\bfr/\partial t$:
\begin{equation}
  \label{eq:chainH}
  \varH_1 = \int dz \sum_{x,y}
  \left[ \frac{p_\bfr^2}{2C} + \frac{U_\parallel}{2}\!\left(\pde{\phi_\bfr}{z}{-}\delta_\bfr\right)^2
    - V_0\cos{M\phi_\bfr} \right],
\end{equation}
where for the moment the inter-chain coupling has been omitted.
The first term and the second term correspond to the change of the total electron kinetic energy
due to temporal and spatial fluctuations, respectively,
whereas the third term represents the commensurability energy.
The dimensionless coupling constants $C$, $V_0$ and $U_\parallel$ depend on such detailed
microscopic structure of the system as the density of states at the Fermi level,
the effective electron mass renormalized due to the lattice vibration,
the electron-phonon coupling strength, and the cutoff energy.\cite{Gruner94}

We now consider inter-chain tunneling of electrons between nearest neighboring chains at $(x,y)$ and
at $(x',y')$ on the $x$-$y$ plane, i.e., at the same position $z$;
this gives rise to the interaction of the form $U_\perp \cos(\phi_\bfr-\phi_{\bfr'})$,
where $U_\perp$ is the dimensionless inter-chain coupling constant and higher-order
harmonics have been disregarded.
With this included, the Hamiltonian $\varH$ for the coupled near-commensurate CDW chains reads
\begin{eqnarray}
  \nonumber
  \varH & = & \int dz
  \left\{
    \sum_{x,y}
    \left[ \frac{p_\bfr^2}{2C}
      + \frac{U_\parallel}{2}\!\left(\pde{\phi_\bfr}{z}{-}\delta_\bfr\right)^2
      - V_0\cos{M\phi_\bfr} \right] \right.\\
  \label{eq:cdwchains}
  & & \left.\mbox\qquad\qquad
    - \sum_{\nn{xy,x'y'}} U_\perp\cos(\phi_\bfr{-}\phi_{\bfr'})
  \right\},
\end{eqnarray}
where the \textit{position} $\phi_\bfr$ and the conjugate \textit{momentum} $p_\bfr$ are considered
to observe the commutation relation $\left[\phi_\bfr,p_{\bfr'}\right]=i\delta_{\bfr,\bfr'}$,
suggesting the position representation $p_\bfr \dot= -i \partial/\partial\phi_\bfr$.
In the second summation $\nn{xy,x'y'}$ stands for the nearest neighbor pairs on the $x$-$y$ plane
at fixed $z$.  Throughout this work, we set $\hbar\equiv 1$, $c\equiv 1$, and
the Boltzmann constant $k_B \equiv 1$ .

Note that in the presence of the commensurability energy, the misfit cannot be simply gauged away
and introduces frustration to the system.  For the time being we consider the case of
strictly commensurate CDW systems without any misfit.
To investigate the quantum phase transition in this case, driven by quantum fluctuations at zero temperature,
we follow the standard procedure\cite{StandardProcedure} to map a $d$-dimensional quantum system
to a $(d{+}1)$-dimensional classical system and obtain the corresponding effective action.
In \secref{sec:homo}, the system with negligible spatial fluctuations along the chain direction is considered
and the corresponding effective action at zero temperature is mapped onto the 3D \XY model,
where commensurability effects are described by the appropriate symmetry-breaking field.
\Secref{sec:general} discusses the system in the presence of spatial fluctuations.
We first pay attention to the classical limit, where temporal fluctuations may be neglected.
Here the system is intrinsically 3D, and the effective action is again mapped onto the 3D \XY model.
Next, with both spatial and temporal fluctuations considered, the appropriate effective action
at zero temperature is identified with the 4D \XY model.

\subsection{Homogeneous Case\label{sec:homo}}

Although it is in general expected that strong fluctuations are present,
we for the moment assume that spatial fluctuations along each chain are negligible,
which gives a two-dimensional (2D) system without $z$ dependence.
In the absence of the symmetry-breaking field due to commensurability,
\eqnref{eq:cdwchains} reduces to the 2D \XY model with kinetic energy
\begin{equation}
  \label{eq:2dxy}
  \varH = \sum_\bfr \frac{p_\bfr^2}{2C} - \sum_{\nn{\bfr,\bfr'}} U_\perp \cos(\phi_\bfr - \phi_{\bfr'}),
\end{equation}
where the summation in the second term is to be performed over all nearest-neighboring pairs
in the 2D space with $\bfr\equiv(x,y)$.  The Hamiltonian in \eqnref{eq:2dxy} has been studied
in the context of quantum arrays of Josephson junctions.\cite{JJ}
In particular, at zero temperature the 2D quantum system in \eqnref{eq:2dxy} is well known
to map onto a 3D classical system via the standard lore.\cite{StandardProcedure}
Introducing the imaginary time $\tau$ axis and dividing the interval between $\tau=0$ and $\tau=T^{-1}$
into $N$ slices of equal width $\Delta\tau=1/NT$, in the zero temperature limit
$(T\rightarrow0)$ we arrive at the partition function of the anisotropic 3D \XY model
\begin{equation}
  \label{eq:an3dxyZ}
  \varZ =\oint\varD\phi\, \exp\left[ \sum_{\nn{\bfr,\bfr'}} K_{\bfr,\bfr'} \cos(\phi_\bfr{-}\phi_{\bfr'})
\right],
\end{equation}
where $\bfr\equiv(\tau,x,y)$ represents the position in the 3D space, consisting of the (imaginary) time
$\tau$ and the 2D space $(x,y)$.  The anisotropic coupling is defined on each bond:
\begin{equation}
  \nonumber
  K_{\bfr,\bfr'} = \left\{
    \begin{array}{ll}
      \displaystyle C/\Delta\tau & \mbox{for } \bfr' = \bfr\pm\hat\bftau\Delta\tau,\\ [2mm]
      U_\perp\Delta\tau & \mbox{for } \bfr' = \bfr\pm \ux \mbox{  or  }\bfr\pm \uy.
    \end{array}
  \right.
\end{equation}
Strictly speaking, we should keep $\Delta\tau$ infinitesimal. Without affecting the universality,
however, we can rescale the space and time and obtain the partition function of an isotropic 3D \XY model
\begin{equation}
  \varZ = \oint\varD\phi \, e^{-H}
\end{equation}
with the desired effective action
\begin{equation*}
  -H = K \sum_{\nn{\bfr,\bfr'}} \cos(\phi_\bfr-\phi_{\bfr'}),
\end{equation*}
where $K^{-1}\equiv (CU_\perp)^{-1/2}$ measures the amount of quantum fluctuations.

We then accommodate the commensurability effects at zero temperature, taking advantage of the fact that the
symmetry-breaking field term is diagonal in the position basis. Via the same procedure and rescaling as in the
absence
of the symmetry-breaking field, the effective action for the 3D \XY model obtains
\begin{equation}
  \label{eq:hH}
  -H = K \sum_{\nn{\bfr,\bfr'}} \cos(\phi_\bfr-\phi_{\bfr'})
  + h \sum_\bfr \cos M\phi_\bfr,
\end{equation}
where the symmetry breaking field is given by $h\equiv V_0\Delta\tau = V_0\sqrt{C/U_\perp}$.

\subsection{General Case\label{sec:general}}

We now turn to the system in which spatial fluctuations in the $z$ direction are present
(but still without the misfit), and first consider the case of only spatial fluctuations,
with temporal fluctuations negligible.  This corresponds to the classical limit in the sense that
the momentum and the position decouple and the momentum part, which can be integrated out,
does not affect the relevant physics.  In this intrinsically 3D case, it is revealing to
resort to the discrete formulation and replace the integration $\int dz$ by the summation
$\sum_z \Delta z$ with sufficiently small $\Delta z$.  Regarding spatial fluctuation-dependent
energy as the continuum form of the cosine interaction in the discrete representation,
we obtain from \eqnref{eq:cdwchains} the effective Hamiltonian for the 3D \XY model:
\begin{equation}
  \label{eq:sH}
  -\varH = \sum_{\nn{\bfr,\bfr'}} K_{\bfr,\bfr'} \cos(\phi_\bfr{-}\phi_{\bfr'}) + h\sum_\bfr \cos M\phi_\bfr,
\end{equation}
where $\bfr\equiv(x,y,z)$ represents lattice sites in the 3D space, and the coupling strength and the
symmetry-breaking field are given by
\begin{eqnarray*}
  K_{\bfr,\bfr'} & \equiv &
  \left\{
    \begin{array}{ll}
      U_\perp\Delta z & \mbox{for  } \bfr'=\bfr\pm \ux \mbox{  or  } \bfr\pm \uy \\ [2mm]
      U_\parallel/\Delta z & \mbox{for  } \bfr'=\bfr\pm \uz\Delta z\\
    \end{array}\right. \\
  h & \equiv & V_0\Delta z.
\end{eqnarray*}
Interestingly enough, \eqnsref{eq:hH} and (\ref{eq:sH}) show that both the (zero-temperature)
quantum phase transition in the absence of spatial fluctuations and the (finite-temperature)
classical phase transition in the absence of temporal fluctuations are described by the same 3D \XY model
under symmetry breaking fields.  Quantum fluctuations in the former play the role of thermal fluctuations
in the latter, tending to restore symmetry.

We finally consider the general case, where both spatial and temporal fluctuations are significant.
Adding the kinetic energy term to the 3D effective action in \eqnref{eq:sH} and following the same procedure
as that in \secref{sec:homo} at zero temperature, we obtain the 4D \XY model with the effective action
\begin{equation}
  \label{eq:4dxy}
  -H = \sum_{\nn{i,i'}} \varK_{i,i'} \cos(\phi_i-\phi_{i'}) + h \sum_i \cos M\phi_i,
\end{equation}
where $i\equiv(\tau,x,y,z)$ denotes 4D space-time lattice sites and the anisotropic coupling
and the symmetry-breaking field are given by
\begin{eqnarray*}
  \varK_{i,i'} &\equiv&
  \left\{
    \begin{array}{ll}
      \displaystyle \frac{\Delta\tau}{\Delta z} U_\parallel & \mbox{ for  } i' = i\pm\uz\Delta z\\ [2mm]
      \displaystyle \frac{C}{\Delta z\Delta\tau} & \mbox{ for  } i' = i\pm\hat\bftau\Delta\tau\\ [4mm]
      \Delta z\Delta\tau U_\perp & \mbox{ for other nearest neighbors}
    \end{array}
  \right. \\
  h &\equiv& V_0\Delta z \Delta\tau
\end{eqnarray*}

In this manner, the coupled near-commensurate CDW systems (without misfit) can be described
by the appropriate \XY models under symmetry-breaking fields.  Depending on whether spatial
and/or temporal fluctuations are present, the effective dimension of the system is determined to be
three or four: It is four if both fluctuations are significant at zero temperature and three otherwise.
The commensurability effects are described by the symmetry-breaking field,
which introduces Z$_M$ symmetry to the system.  Such a symmetry-breaking field affects the
ground-state symmetry and is expected to be relevant in the sense of the renormalization group (RG)
theory.  In the limit $M\rightarrow\infty$, however, the Z$_M$ symmetry is hardly distinguishable
from the underlying U(1) symmetry in the \XY model.  Therefore we expect the phase transition
to be crucially dependent upon the commensurability factor $M$, and devote the next two sections
to the investigation of the phase transitions in the 3D and 4D \XY models under symmetry-breaking fields.

%
\section{3D \XY Model under Symmetry-Breaking Field\label{sec:3dxy}}

In this section we investigate the phase transition in the 3D \XY model whose effective action is
given by \eqnref{eq:hH}.
In the absence of the symmetry-breaking field ($h=0$), the 3D \XY model has been studied both
analytically\cite{VortexLoop,Williams88,Shenoy89} and numerically,\cite{3DXYNumericalStudy}
revealing that vortex loops do exist and proliferate at the phase transition.
Accordingly, the topological scaling idea has been extended to the 3D transition with conventional
long-range order, with the scaling procedure of Ref.~\onlinecite{VortexScaling} generalized
appropriately for 3D directed loops.\cite{Williams88,Shenoy89}
One may then be tempted to extend the study of the 2D model in Ref.~\onlinecite{Jose77}
to incorporate the symmetry-breaking term in the 3D \XY model,
and combine recursion relations for the vortex fugacity $y_0$ and the field perturbation $y_h$
according to the duality relation between them.
However, in contrast to the 2D case, the geometric scaling of the coupling constant in three dimensions
results in nonzero values of $y_0$ and $y_h$ at the fixed point.\cite{Shenoy89}

On the other hand, the $\epsilon$-expansion\cite{Wilson72} is well known to provide a mathematical
formalism for calculating critical exponents of the O($n$) model near four spatial dimensions,
allowing classification of the universality class of the system.
Here we extend the $\epsilon$-expansion approach to incorporate the effects of the symmetry-breaking field.
The $\epsilon$-expansion turns out to be useful only for small commensurability factor $M$;
we thus supplement its limitations with the Monte Carlo numerical method.
In \secref{sec:eexp}, the symmetry-breaking perturbation is treated within the $\epsilon$-expansion
and \secref{sec:3dxymc} is devoted to the Monte Carlo simulations of the 3D \XY model
under symmetry-breaking fields.

%
\subsection{$\epsilon$-Expansion\label{sec:eexp}}

We consider the $\epsilon$-expansion for the 3D \XY model, extending the original formulation for the O($n$)
model\cite{Wilson72} to incorporate the symmetry-breaking field.
To begin with, we employ the two-component continuous local variable or spin
$\bfS_\bfr=(S_\bfr^x,S_\bfr^y)$ at each 3D lattice site $\bfr$, the total number of which is denoted by $N$. 
The constraint that each spin has
the unit magnitude is relaxed by the additional weight factor introduced to the partition function:
\begin{equation}
  \varZ = \left(\prod_\bfr \int_{-\infty}^\infty d\bfS_\bfr\right) e^{-[H+W(\bfS_\bfr)]}
\end{equation}
with the weight factor ($u>0$)
\begin{equation*}
  \exp[-W(\bfS)]=\exp\left[-\frac{1}{2}bS^2-uS^4\right],
\end{equation*}
which has been expanded up to $\varO(S^4)$. At the physical \XY fixed point the sixth-order term is less
relevant in dimension $d=4-\epsilon$ than $d=4$.\cite{Cardy96} With the identification
$S_\bfr^x = \cos\phi_\bfr$ and $S_\bfr^y=\sin\phi_\bfr$, the original action in \eqnref{eq:hH} is
written in terms of the continuous spin $\bfS_\bfr$:
\begin{eqnarray}
  \nonumber
  -H & = & \sum_{\bfr,\bfr'} K \bfS_\bfr\cdot\bfS_{\bfr'}
  {+} \frac{h}{2} \sum_\bfr \left[ (2S^x_\bfr)^M {-} \frac{M}{1!}(2S^x_\bfr)^{M-2} \right.\\
  \label{eq:eeH}
  & & \left.\mbox{} \qquad\qquad\quad{+} \frac{M(M{-}3)}{2!}(2S^x_\bfr)^{M-4} {-} \cdots \right].
\end{eqnarray}
Note that for $M<4$ the order of the action does not exceed $\varO(S^4)$,
leading to the Ginzburg-Landau-Wilson (GLW) effective Hamiltonian $\varH=H+W$ to $\varO(S^4)$.

We first examine the case $M=2$. In the momentum space representation, keeping only relevant terms to
$\varO(q^2)$ and scaling the spin variable according to $\bfS_\bfq = (Ka^{d+2})^{-1/2}\bfsigma_\bfq$,
we express the GLW Hamiltonian as
\begin{eqnarray}
  \nonumber
  -\varH & = & -\half \int_\bfq (r_x{+}q^2)\sigma^x_\bfq\sigma^x_{-\bfq}
  -\half \int_\bfq (r_y{+}q^2)\sigma^y_\bfq\sigma^y_{-\bfq} \\
  \label{eq:fa}
  & & \quad\mbox{} + V_1\int_\bfq \int_{\bfq'} \int_{\bfq''}
  \sigma^x_\bfq\sigma^x_{\bfq'}\sigma^x_{\bfq''}\sigma^x_{-\bfq-\bfq'-\bfq''} \\
  \nonumber
  & & \quad\mbox{} + 2V_2\int_\bfq \int_{\bfq'} \int_{\bfq''}
  \sigma^x_\bfq\sigma^x_{\bfq'}\sigma^y_{\bfq''}\sigma^y_{-\bfq-\bfq'-\bfq''} \\
  \nonumber
  & & \quad\mbox{} +V_3\int_\bfq \int_{\bfq'} \int_{\bfq''}
  \sigma^y_\bfq\sigma^y_{\bfq'}\sigma^y_{\bfq''}\sigma^y_{-\bfq-\bfq'-\bfq''},
\end{eqnarray}
where the coefficients are given by
\begin{align*}
  & r_x =\frac{1}{Ka^2}(b-dK-2h) \\
  & r_y =\frac{1}{Ka^2}(b-dK) \\
  & V_1 = V_2= V_3= -u < 0
\end{align*}
and $\int_\bfq \equiv \int d\bfq/(2\pi)^d = (1/Na^d)\sum_\bfq$ with the lattice constant $a$
restored for clarity. Here it is shown that the $M=2$
symmetry-breaking field gives rise to anisotropy in the quadratic term, making $r_x$ less than $r_y$.
Following the standard procedure, we obtain the recursion relations to the leading order in $\epsilon\equiv4-d$:
\begin{eqnarray}
  \nonumber
  \pde{r_x}{l} &=& 2r_x - \frac{12C}{1+r_x}V_1 - \frac{4C}{1+r_y}V_2 \\
  \nonumber
  \pde{r_y}{l} &=& 2r_y - \frac{4C}{1+r_x}V_2 - \frac{12C}{1+r_y}V_3 \\
  \label{eq:eerr}
  \pde{V_1}{l} &=& \epsilon V_1 + \frac{36C}{(1+r_x)^2}V_1^2 + \frac{4C}{(1+r_y)^2}V_2^2 \\
  \nonumber
  \pde{V_2}{l} &=& \epsilon V_2 + \frac{16C}{(1+r_x)(1+r_y)}V_2^2 \\
  \nonumber
  & & \mbox{} + \frac{12C}{(1+r_x)(1+r_y)}V_1V_2 + \frac{12C}{(1+r_x)(1+r_y)}V_2V_3 \\
  \nonumber
  \pde{V_3}{l} &=& \epsilon V_3 + \frac{36C}{(1+r_x)^2}V_3^2 + \frac{4C}{(1+r_y)^2}V_2^2
\end{eqnarray}
with the spatial scale factor $l$ and $C\equiv2^{1-d}\pi^{d/2}/(d/2{-}1)!$.

In the absence of the symmetry-breaking field ($h=0$), the parameter space reduces to the 2D space
$(r,V)$ since $r_x=r_y\equiv r$ and $V_1=V_2=V_3\equiv V$.
In this case the nontrivial fixed point of the recursion relation in \eqnref{eq:eerr} is
simply the \XY fixed point, given by
\begin{equation}
  r^* = -\frac{\epsilon}{5},\quad V^* = -\frac{\epsilon}{40C}
\end{equation}
to $\varO(\epsilon)$.  As the symmetry-breaking field is turned on, however, we have $r_x<r_y$
at the initial locus and need to examine the RG flow in the full five-dimensional (5D) parameter space
$(r_x,r_y,V_1,V_2,V_3)$.  To investigate the stability of the \XY fixed point in the 5D parameter space,
we linearize \eqnref{eq:eerr} about this point and obtain a stability matrix which yields five eigenvalues
together with corresponding scaling exponents:
\begin{equation}
  \label{eq:ev}
  \begin{array}{c}
    \displaystyle y_1=2-\frac{\epsilon}{5},\quad y_2=2-\frac{2\epsilon}{5},\quad \\
    \displaystyle y_3 = -\epsilon,\quad y_4 = -\frac{\epsilon}{5},\quad y_5 = -\frac{4\epsilon}{5}.
  \end{array}
\end{equation}
Since the exponents $y_3$, $y_4$, and $y_5$, which correspond to the three eigenvectors spanning
the 3D subspace $(V_1,V_2,V_3)$, are all strictly negative, it is concluded that all $V_i$'s (for $i=1,2,3$)
are irrelevant at the \XY fixed point.  In contrast, both $y_1$ and $y_2$ are positive, indicating that
the \XY fixed point is unstable in the 2D subspace $(r_x,r_y)$. The initial locus,
which is kept off the \XY fixed point by the symmetry-breaking field, should
flow far off the \XY fixed point. Instead it is expected to flow toward the Ising fixed point located at
\begin{equation}
  r_x^* = -\frac{\epsilon}{6},\quad  r_y^* = \infty, \quad  V_1^* = -\frac{\epsilon}{36C},\quad  V_2^*=V_3^*=0.
\end{equation}
At this point it is more appropriate to take $r^{-1}_y$ as the scaling field, giving the scaling exponents
\begin{equation}
  y_{r_x} = 2-\frac{\epsilon}{3},\qquad y_{r_y^{-1}} = -2.
\end{equation}
Thus the initial locus with $r_x<r_y$ indeed flows toward the Ising fixed point along the stable
$r_y$ direction.  In this manner, the symmetry-breaking field for $M=2$ introduces anisotropy in
the quadratic term, making the \XY fixed point unstable and generating an Ising or two-state clock fixed point.
Revealed accordingly is a single second-order transition between two-state clock order and disorder.

We next turn to the case $M=3$, where \eqnref{eq:eeH} shows that terms of linear and third order
in the $x$-component of the spin come into play; power counting suggests that these fields are
relevant near $d=4$.  Owing to the anisotropy associated with the absence of $S^y$ terms, in particular,
the cubic term $({S^x})^3$ here may not be removed by mere shift and is not redundant,
in contrast to the case of the Ising model. The symmetry-breaking field is thus expected
to drive the transition between the disordered phase and the three-state clock ordered phase.

Finally the symmetry-breaking field for $M=4$ introduces anisotropy in the quartic term:
\begin{equation}
  \label{eq:M4}
  \begin{array}{l}
    V_1=-u + 8a^{d-4}K^{-2}h \\
    V_2=V_3=-u\ .
  \end{array}
\end{equation}
This leads the action in \eqnref{eq:fa} to be unstable for sufficiently large values of the field $h$,
making it necessary to consider higher-order terms in the weight function $W(\bfS)$ for stability.
Unfortunately, such calculation of higher cumulants is quite a formidable job, and it is
very difficult to extend the $\epsilon$-expansion to the case of a high commensurability factor.

%
\subsection{Monte Carlo Simulations\label{sec:3dxymc}}

This section presents the Monte Carlo study of the 3D \XY model in symmetry-breaking fields.
To estimate the critical temperatures and determine critical exponents, we have performed Monte Carlo
simulations at ``temperatures'' (i.e., quantum or temporal fluctuations) ranging from
$K^{-1}=0.5$ to $K^{-1}=5$ on lattices of linear size $L=4$ up to $L=32$
for several commensurability factors and field strengths.
Measured in simulations are the order parameter $m$ and the susceptibility $\chi$ defined to be
\begin{equation}
  \begin{array}{rcl}
    m & \equiv & \displaystyle \avg{\left|\rcp{N}\sum_\bfr e^{i\phi_\bfr}\right|} \\ [4mm]
    \chi & \equiv & \displaystyle \avg{\left|\rcp{N}\sum_\bfr e^{i\phi_\bfr}\right|^2}
    - \avg{\left|\rcp{N}\sum_\bfr e^{i\phi_\bfr}\right|}^2,
  \end{array}
\end{equation}
where $N$ is the number of sites.
We have employed the multiple histogram method\cite{MonteCarlo} to interpolate the
quantities calculated sparsely in a given range of the temperature to any temperature inside the range.
It not only saves a great deal of computing time but also provides the values of quantities
at arbitrary temperatures for finite-size scaling, resulting in critical exponents of better accuracy.
In fact we have obtained the best collapse of the scaling function such as
\begin{equation}
\tilde m (L^{1/\nu}t) = L^{\beta/\nu} m(t)
\end{equation}
with the reduced temperature $t\equiv T/T_c-1 \equiv K_c/K -1$,
by minimizing the measure of error\cite{MonteCarlo}
\begin{equation}
  \sigma_m^2 \equiv \rcp{x_{max}{-}x_{min}} \int_{x_{min}}^{x_{max}} \!\!\!dx 
     \left[\left\langle\!\left\langle\tilde m^2(x)\right\rangle\!\right\rangle
    - \left\langle\!\left\langle\tilde m(x)\right\rangle\!\right\rangle^2 \right]
\end{equation}
over the critical temperature and exponents.
Here $\left\langle\!\left\langle\cdots\right\rangle\!\right\rangle$ stands for
the average over the lattice size and similar relations for the susceptibility have 
also been considered. 
We have taken the range $[x_{min}, x_{max}] = [-\Delta x, \Delta x]$ and attempted
a series of collapses as $\Delta x$ is diminished.  The result has then been extrapolated
to the limit $\Delta x\rightarrow 0$.  

\begin{figure}
  \centering
  \includegraphics[width=8cm]{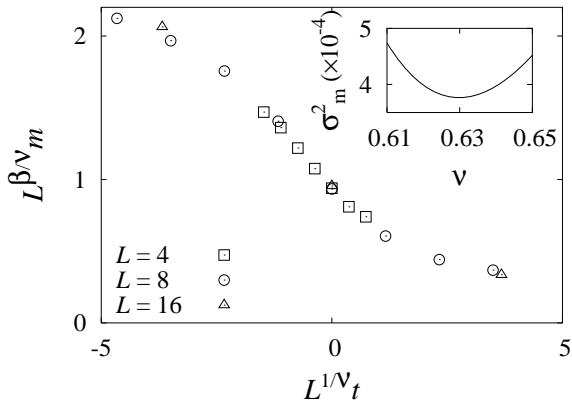}
  \caption{Data collapse of the order parameter $m$ for three different values of the system size $L$, 
    with the best-estimated critical temperature and exponents given in the text. 
    The inset shows the dependence of the error $\sigma_m^2$ on the value of $\nu$, 
    with $T_c$ and $\beta$ fixed at their best estimated values.  
    The data have been taken for $M=2$ and at $h=2.7$; typical error bars are not larger than the symbol sizes.}
  \label{fig:3dxys}
\end{figure}
\Figref{fig:3dxys} presents typical collapse of the
scaling function with the best-estimated critical temperature $T_c$ and exponents $\beta$ and $\nu$; 
nice collapse behavior can be observed near the critical temperature $(t=0)$. 
The inset in \figref{fig:3dxys} discloses how the corresponding error $\sigma_m^2$ depends on
the value of the exponent $\nu$.  We have thus estimated the error in the obtained critical exponent 
from the standard deviation of the minimizing values over the sample ensemble.
In this way, the phase transition for $M=2$ is found to be of the second order with the critical exponents
$\nu=0.63\pm0.01$ and $\beta=0.34\pm0.04$ for $h = 1.2$ and
$\nu=0.627\pm0.004$ and $\beta=0.32\pm0.01$ for $h = 2.7$.
These results coincide perfectly with the known critical exponents for the 3D Ising model:
$\nu=0.630$ and $\beta=0.324$, thus demonstrating the validity of the $\epsilon$-expansion
analysis in \secref{sec:eexp}.
Here it is of interest that the same universality class was also reported in the quantum Monte Carlo study 
of a tight-binding model of spinless fermion chains coupled by intra- and inter-chain Coulomb
interactions.\cite{Scalapino86}  The tight-binding model analysis, though being restricted to the $M=2$ case, 
could incorporate amplitude fluctuations of the CDW order parameter directly. 
Indeed the accordance between the two analyses manifests that fluctuations in the phase (rather than in the 
amplitude) mostly determine the nature of the transition. 

\begin{figure}
  \centering
  \includegraphics[width=8cm]{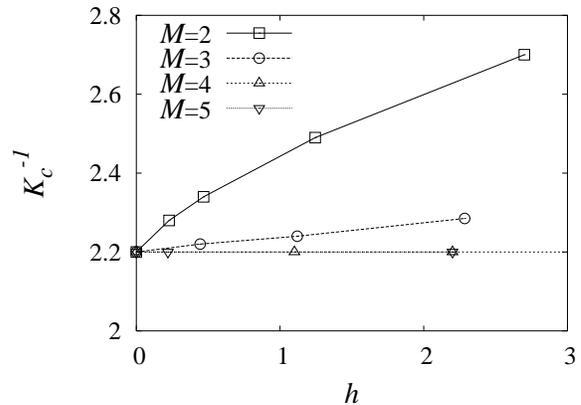}
  \caption{Critical temperature $K_c ^{-1}$ in the 3D \XY model under symmetry-breaking fields
   versus the symmetry-breaking field strength $h$, for the commensurability factor $M=2, 3, 4$, and 5.
   Typical error bars are smaller than the symbol size and lines are merely guides to eyes.}
  \label{fig:3dxytc}
\end{figure}

For $M=3$, on the other hand, we have found no evidence for the first-order transition,
and obtained the estimation $\nu=0.596\pm0.007$ and $\beta=0.3\pm0.01$;
this is to be compared with the long-standing controversy as to whether the transition
in the 3D three-state Potts model is of the first order or continuous.\cite{PottsModel}
It is also of interest that the critical behavior for $M\ge4$ is similar to that of the 3D \XY model
without any symmetry-breaking field.  \Figref{fig:3dxytc} shows the critical temperature
(i.e., the critical fluctuation strength) $K_c$ versus the field strength $h$ for $2\le M\le5$.
It is observed that for $M\le3$ the critical temperature increases with the field strength
while for $M\ge4$ the commensurability energy appears to have no effect on the critical temperature,
in the range probed.  Such critical behavior for $M\ge4$, similar to that of the ordinary 3D
\XY model, reflects that, as noted in \secref{sec:ea}, the Z$_M$ symmetry for large $M$ is
indiscernible from the U(1) symmetry underlying in the \XY model.
%

%
%

\section{4D \XY Model under Symmetry-Breaking Field\label{sec:4dxy}}

We now study the phase transition in the 4D \XY model, described by the effective action
in \eqnref{eq:4dxy}, through the use of the mean-field approximation.
Here the mean-field approximation, developed for superconducting arrays in
applied magnetic fields,\cite{Shih83} is expected to be accurate since the upper critical dimension
of the \XY model is given by $d_u=4$.  With the space and time rescaled appropriately,
the (mean-field) self-consistent equation reads
\begin{equation}
  \label{eq:ssc}
  \avg{e^{i\phi}} = \varZ_{MF}^{-1} \int_0^{2\pi} d\phi\, e^{i\phi} e^{-H_{MF}},
\end{equation}
where $\varZ_{MF}$ is the partition function
\begin{equation*}
  \varZ_{MF} = \int_0^{2\pi} d\phi\, e^{-H_{MF}}
\end{equation*}
corresponding to the mean-field action
\begin{equation}
  -H_{MF} = 4K(\avg{\cos\phi} \cos\phi + \avg{\sin\phi}) \sin\phi + h \cos M\phi
\end{equation}
with the rescaled coupling constant $K\equiv \sqrt{CU_\perp}$ and field $h\equiv V_0 \sqrt{C/U_\perp}$.

In the absence of the symmetry-breaking field, the self-consistent equation leads to the equation of state
for the order parameter $m\equiv\sqrt{\avg{\cos\phi}^2+\avg{\sin\phi}^2}$:
\begin{equation}
  m = \frac{I_1(4Km)}{I_0(4Km)},
\end{equation}
where $I_n$ is the $n$th modified Bessel function.  The system is in the disordered phase for
$K<K_c(h{=}0)=1/2$ characterized by $m=0$;
beyond $K_c(h{=}0)$ the ordered phase with $m\ne0$ is favored.
Similarly, the $M=1$ case can be analyzed easily, where arbitrarily small but positive $h$
results in $\avg{\cos\phi}>0$ and $\avg{\sin\phi}=0$.  The equation of state is thus
\begin{equation}
  \avg{\cos\phi} = \frac{I_1(4K\avg{\cos\phi}+h)}{I_0(4K\avg{\cos\phi}+h)}.
\end{equation}

For larger values of $M$, \eqnref{eq:ssc} may be solved numerically with the parameters $K$ and $h$ varied.
It is found that there exists a field-dependent transition coupling strength $K_c(h)$ below which
\eqnref{eq:ssc} bears only the null solution $\avg{\cos\phi} = \avg{\sin\phi} = 0$.
As $K$ is increased beyond $K_c(h)$, nonzero stable solutions emerge.
Due to the Z$_M$ symmetry, i.e., the invariance of the action under the shift in the angle by $2\pi n/M$ for
any integer $n$, the solution of \eqnref{eq:ssc} can be expressed in terms of the $M$-state clock order
parameter $m(K,h)$:
\begin{equation}
  \avg{e^{i\phi}} = m(K,h) e^{2\pi i n/M}
\end{equation}
with integer $n=0,1,\ldots,M-1$. The critical coupling strength $K_c(h)$ is defined to be the largest value
of $K$ satisfying $m(K,h)=0$.

\begin{figure}
  \centering
  \includegraphics[width=8cm]{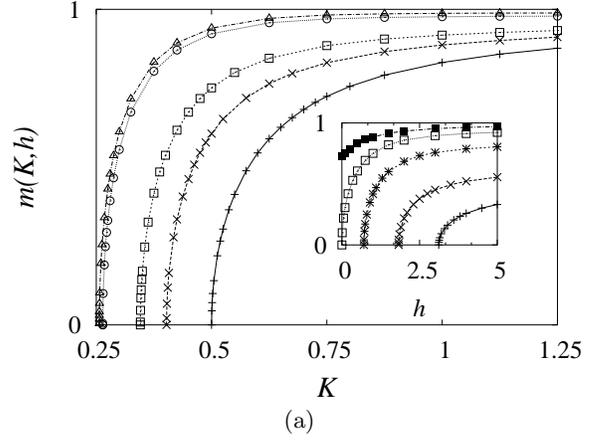}\\ (a)\\[1cm]
  \includegraphics[width=8cm]{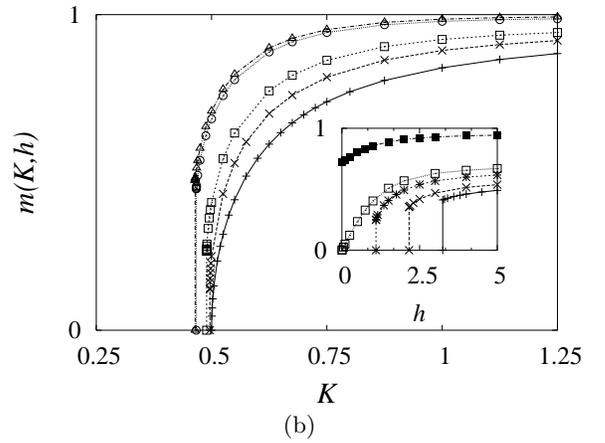}\\ (b)\\[1cm]
  \includegraphics[width=8cm]{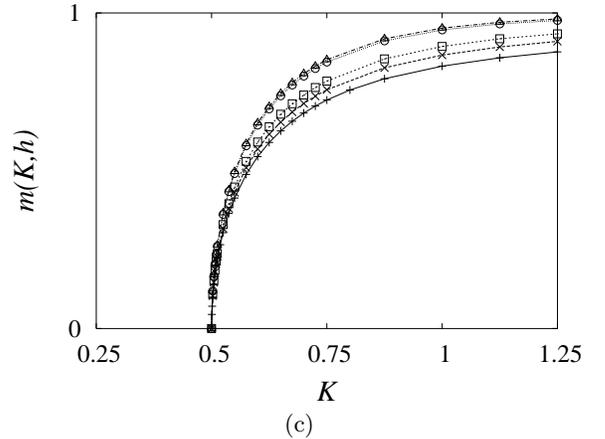}\\ (c)
  \caption{Order parameter $m$ as a function of the coupling $K$ and the field strength $h$ for (a) $M=2$,
   (b) $M=3$, and (c) $M=4$.  Main graphs show $m$ versus $K$ for $h = 0(+), 0.5(\times),
    1(\square), 5(\bigcirc)$, and $10(\bigtriangleup)$ from below.  Insets display $m(K,h)$ versus $h$ for
    various values of $K$: From below, (a) $K = 0.275, 0.3, 0.375, 0.5$, and $0.75$;
   (b) $K = 0.47, 0.475, 0.4875, 0.5$, and $0.75$.  The overall behavior of the order parameter for $M>4$
    is the same as that for $M=4$.}
  \label{fig:mforder}
\end{figure}
\Figref{fig:mforder} shows how the order parameter $m$ depends on the parameters $K$ and $h$
for different values of $M$.  It is observed that $m$ in general increases rapidly from zero as $K$ exceeds $K_c$
and then saturates to unity, with the increase more rapid for larger $h$.
While for $M=2$ and $M\ge4$ the transition is found to be of the second order,
numerical results for $M=3$ indicate a first-order transition, which is expected from the appearance
of the third-order term in the $\epsilon$-expansion.  For $M\le3$, the critical coupling strength $K_c(h)$,
starting from $K_c(h{=}0)=1/2$, decreases with $h$ almost exponentially to the asymptotic values,
0.250 and 0.462 for $M=2$ and for $M=3$, respectively.  For $M\ge4$, on the other hand,
the transition point does not depend on the strength of the symmetry-breaking field,
giving $K_c=1/2$ regardless of $h$.  Again manifested is the crucial role of commensurability
in the phase transition.

\begin{figure}
  \centering
  \includegraphics[width=8cm]{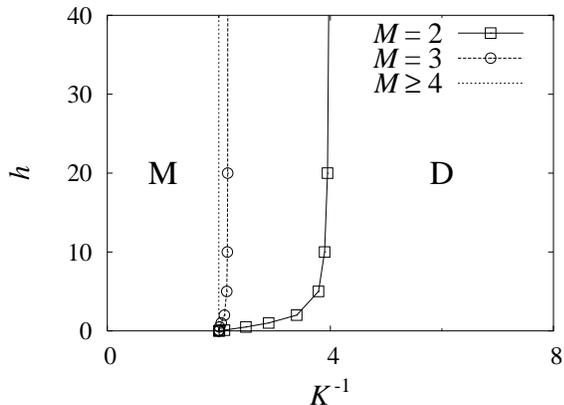}
  \caption{Phase diagram for the 4D \XY model under the symmetry-breaking field, displaying the boundaries
    between the $M$-state clock order phase (M) and the disordered phase (D) for different values of $M$.
    Lines for $M = 2$ and $3$ are merely guides to eyes.}
  \label{fig:mfpd}
\end{figure}

The resulting phase diagram for the 4D \XY model under the symmetry-breaking field is displayed in
\figref{fig:mfpd}, where the boundary $K_c^{-1}$ versus $h$ is plotted for different values of $M$.
When $K^{-1}<K_c^{-1}$, the Z$_M$ symmetry as well as the U(1) symmetry is broken and identified
is the $M$-state clock ordered phase, with one of the minima of $-h\cos M\phi$ favored.
When $K^{-1}>K_c^{-1}$, unlike the U(1) symmetry broken explicitly for $h\ne0$,
the Z$_M$ symmetry remains unbroken, leading to the disordered phase.

%
\section{Commensurate-Incommensurate Transition\label{sec:cic}}

Up to the present, we have assumed the absence of misfit, so that only the transition between
the disordered phase and the $M$-state clock ordered phase, where commensurate CDWs are developed,
has been considered in several idealized cases described in \secref{sec:ea}.
However, the misfit, being a key ingredient to bring about the commensurate-incommensurate transition
in a near-commensurate CDW chain, must be taken into consideration in understanding various transitions
in the coupled CDW system.

It is well known from the study of the 1D Frenkel-Kontorova model\cite{Choi00} at zero temperature that
a single near-commensurate CDW changes from the commensurate state to the incommensurate state
when the misfit exceeds a critical value depending on the commensurability energy.  At finite temperatures
the 1D CDW system is always in the incommensurate state.\cite{Sahni81}
On the other hand, in the coupled CDW system, interactions between the CDW chains may alter the nature
of the transition.  First of all, the effective dimensions of the system grow to three, giving rise to
the persistence of long-range order, as observed in the previous sections.  Moreover, the CI transition itself
can also be affected by the inter-chain coupling in that the interactions favor either the commensurate state
or the incommensurate state, as explained below.  To investigate the phase transition of the coupled CDW
system, we consider the Hamiltonian in \eqnref{eq:cdwchains} without temporal fluctuations,
which makes the problem simpler and allows to focus on the static properties only.

At zero temperature the problem is rather simple to solve.  The inter-chain coupling term in
\eqnref{eq:cdwchains} reaches the minimum when all the phases $\phi_\bfr$ in the $xy$-plane
become equal to each other.  All the CDW chains, therefore, follow the same phase configurations
determined by the 1D Frenkel-Kontorova model, thus undergo the CI transition at the critical misfit
given by\cite{Choi00}
\begin{equation}
  \delta_c = \frac{4}{\pi} \sqrt{\frac{V_0}{U_\parallel}},
\end{equation}
regardless of the inter-chain coupling strength $U_\perp$.

The system at finite temperatures is examined by means of the Monte Carlo method.  For this we discretize
the $z$-axis in \eqnref{eq:cdwchains} as in \secref{sec:general}, and obtain the lattice Hamiltonian
\begin{eqnarray}
  \nonumber
  \varH & = &
  \sum_\bfr \frac{U_\parallel}{2} \!\left(\phi_{\bfr+\bfz}{-}\phi_\bfr{-}\delta\right)^2
  - \sum_\bfr V_0\cos{M\phi_\bfr} \\
  & & \mbox{}
  - \sum_z \sum_{\nn{xy,x'y'}} U_\perp\cos(\phi_\bfr{-}\phi_{\bfr'}),
\end{eqnarray}
where $\Delta z$ has been absorbed into the coupling constants.  Since the dimension along the CDW chain
is usually longer than the lateral dimensions, the lattice size $L_z$ along the $z$-axis is kept to be two times
larger than the other linear sizes in our simulations, and accordingly the inter-chain coupling $U_\perp$ is
restricted to be smaller than $U_\parallel$ to avoid inessential finite-size effects.
We further set the parameters to $U_\parallel=1$ and $V_0=0.2$ and sweep the inter-chain coupling
strength from 0.01 to 0.5 and the misfit $\delta$, which is assumed to be uniform throughout the system,
from zero to 0.8 beyond the critical misfit $\delta_c \,(=0.569)$.

The order-disorder transition is described conveniently by the (incommensurate) order parameter
defined to be
\begin{equation}
  m \equiv \avg{\left|\rcp{N}\sum_\bfr e^{i(\phi_\bfr - \delta z)}\right|}
\end{equation}
and its susceptibility $\chi$.  Note that in the thermodynamic limit this order parameter vanishes
not only in the disordered phase but also in the commensurate phase;
in the system of finite size it may remain non-zero even in the perfectly commensurate state.
We also compute the soliton density $\rho_s$,\cite{Topic90} i.e., the soliton number per site per chain,
which characterizes the CI transition.
For the detection of any hysteresis, these two physical quantities have been measured in two different ways:
in the cooling-down process with randomized initial phase configurations and in the heating-up process
starting from the zero-temperature ground state.  In each process the system has been equilibrated
sufficiently while the temperature is varied gradually.

\begin{figure}[!t]
  \centering
  \includegraphics[width=8cm]{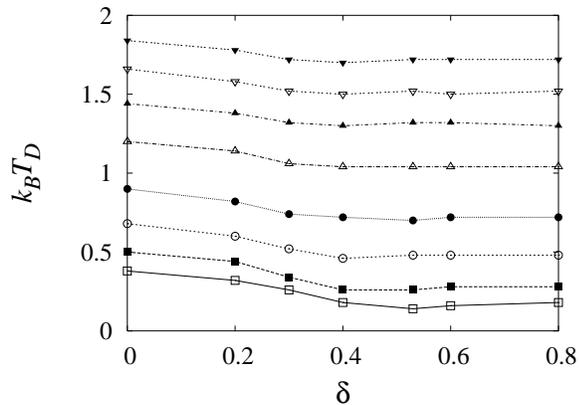}
  \caption{Behavior of the order-disorder transition temperature $T_D$ with the misfit $\delta$
    in the $M=2$ coupled CDW system for $U_\parallel=1$ and $V_0=0.2$.
    Each symbol corresponds to a different value of the inter-chain coupling strength:
    $U_\perp = 0.01(\square), 0.02(\blacksquare), 0.05(\circ), 0.1(\bullet), 0.2(\triangle),
    0.3(\blacktriangle), 0.4(\triangledown)$, and $0.5(\blacktriangledown)$.
    Typical error bars are smaller than the symbol sizes and lines are merely guides to eyes.}
  \label{fig:ct}
\end{figure}
Shown in \figref{fig:ct} is the dependence of the order-disorder transition temperature $T_D$
on the misfit $\delta$ in the $M=2$ CDW system at various inter-chain coupling strengths.
It is observed that the misfit tends to reduce the transition temperature:
The transition temperature first decreases as the misfit is increased from zero, then saturates
when the misfit reaches a value comparable to $\delta_c$.
Here stronger inter-chain coupling, which helps to increase the transition temperature,
in general weakens the effects of the misfit.  Besides the transition temperatures, the critical exponents
also change with the misfit.  For instance, the critical exponents in the system with $U_\perp=0.1$ are
found to be $\nu = 0.79\pm0.02$ and $\gamma=1.33\pm0.01$ at $\delta=0.2$;
$\nu = 0.77\pm0.03$ and $\gamma=1.64\pm0.05$ at $\delta=0.4$.
This suggests that the introduction of the misfit changes the nature of the ordered phase,
making it different from the 3D \XY ordered phase.  It should be identified as the
incommensurate CDW phase, as demonstrated below.

\begin{figure}
  \centering
  \includegraphics[width=8cm]{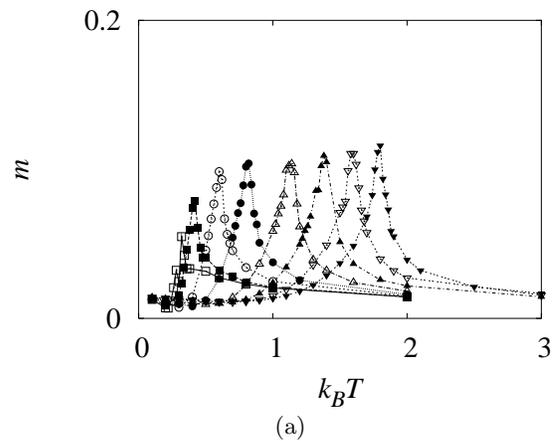}\\ (a)\\[0.5cm]
  \includegraphics[width=8cm]{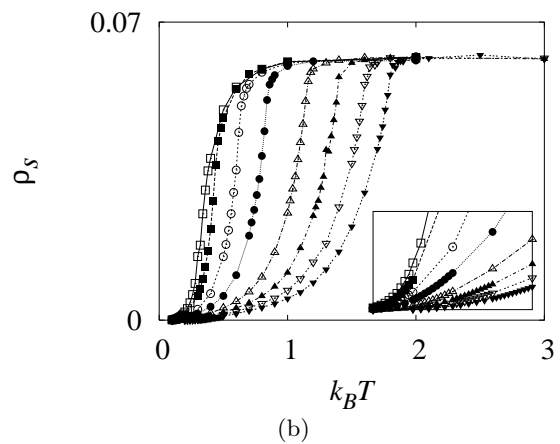}\\ (b)
  \caption{(a) The order parameter $m$ and (b) the soliton density $\rho_s$ versus the temperature $T$
    in the $M=2$ system of misfit $\delta=0.2$ and size $L_z=32$.
    Each symbol corresponds to a different value of the inter-chain coupling strength as listed
    in \figref{fig:ct}.  The inset is an enlarged view of the soliton density in the range $0.1\le T\le0.5$.}
  \label{fig:ccdw_d0.2}
\end{figure}
We now draw our attention to the CI transition occurring at lower temperatures.
In the absence of misfit ($\delta=0$) the soliton density $\rho_s$ is observed to vanish on average
at all temperatures below $T_D$, indicating that the CDWs formed are commensurate even in the
presence of thermal fluctuations.  Such a commensurate phase is destroyed by the introduction
of misfit, even for $\delta<\delta_c$, with the help of thermal fluctuations.
\Figref{fig:ccdw_d0.2} exhibits the behaviors of the order parameter and of the soliton density
as the temperature is varied in the system with misfit $\delta=0.2$.
It is shown that the order parameter reaches its maximum at temperature near $T_D$ and decreases
to zero as the temperature is reduced while the soliton density begins to decrease from its maximum
value at $T\approx T_D$ and vanishes almost to zero at very low temperatures.
Nonzero values of the order parameter and of the soliton density together in the regime
$T\lesssim T_D$ correspond to the formation of incommensurate CDWs.
Thus the order-disorder transition at $T=T_D$ is identified as the transition between
the incommensurate CDW state and the disordered state.
In addition, the vanishing soliton density [see the inset in \figref{fig:ccdw_d0.2}] is a sign of
the commensurate CDW phase; this indicates the presence of a CI transition driven by thermal fluctuations.
The CI transition temperature $T_{IC}$ is defined to be the temperature at which the soliton density
begins to be non-zero (in the thermodynamic limit).
Although our computing ability disallow us to determine the precise value of $T_{IC}$,
\figref{fig:ccdw_d0.2}(b) manifests that $T_{IC}$ increases with the inter-chain coupling strength.
In this sense the inter-chain interactions favor the commensurate state.  In contrast,
it is also observed that the inter-chain interactions prefer the incommensurate state at the misfit
values somewhat larger (but still smaller than $\delta_c$).

\begin{figure*}
  \centering
  \begin{minipage}{8cm}
    \centering
    \includegraphics[width=\textwidth]{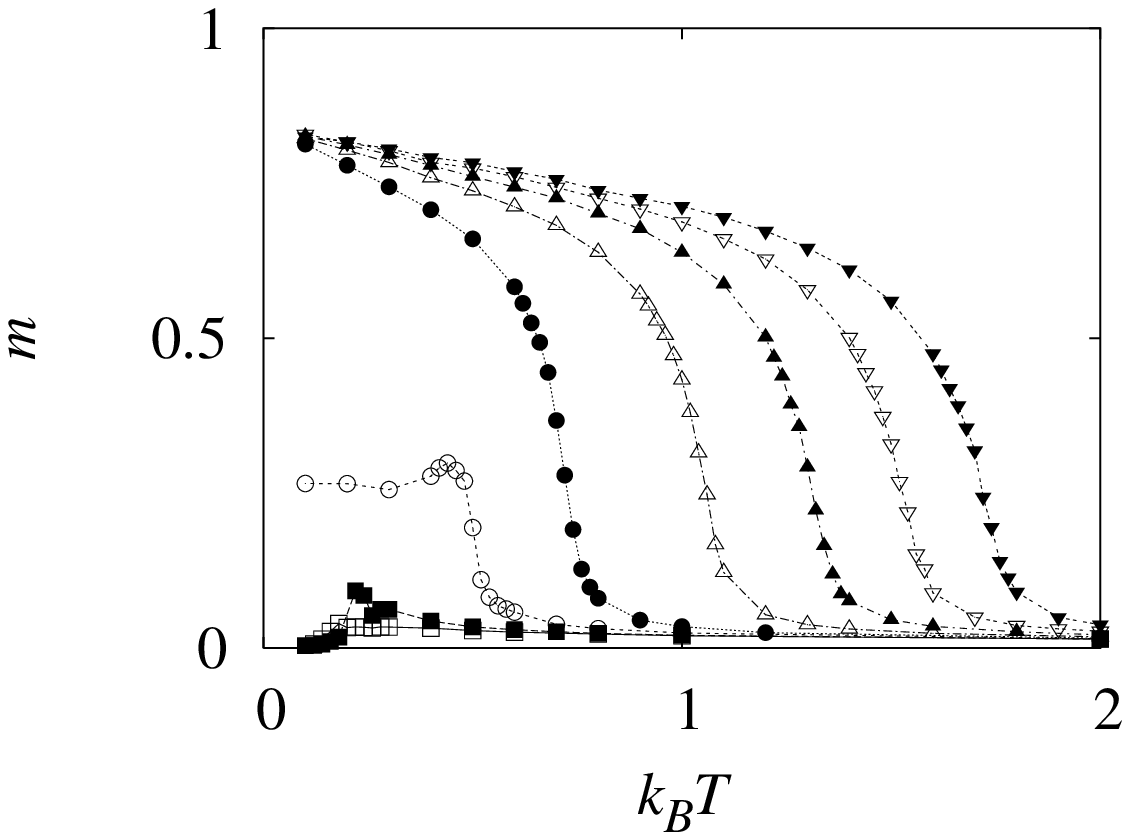}
    (a)
  \end{minipage}
  \begin{minipage}{8cm}
    \centering
    \includegraphics[width=\textwidth]{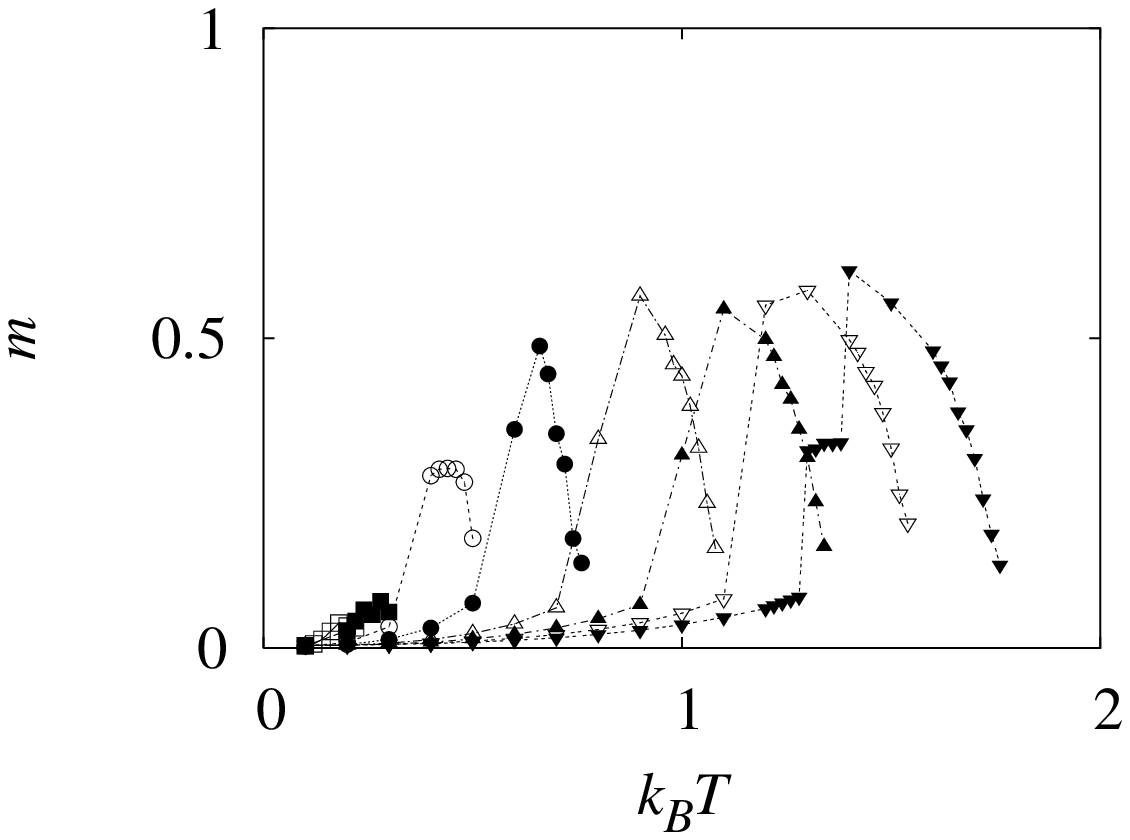}
    (b)
  \end{minipage}\\[0.5cm]
  \begin{minipage}{8cm}
    \centering
    \includegraphics[width=\textwidth]{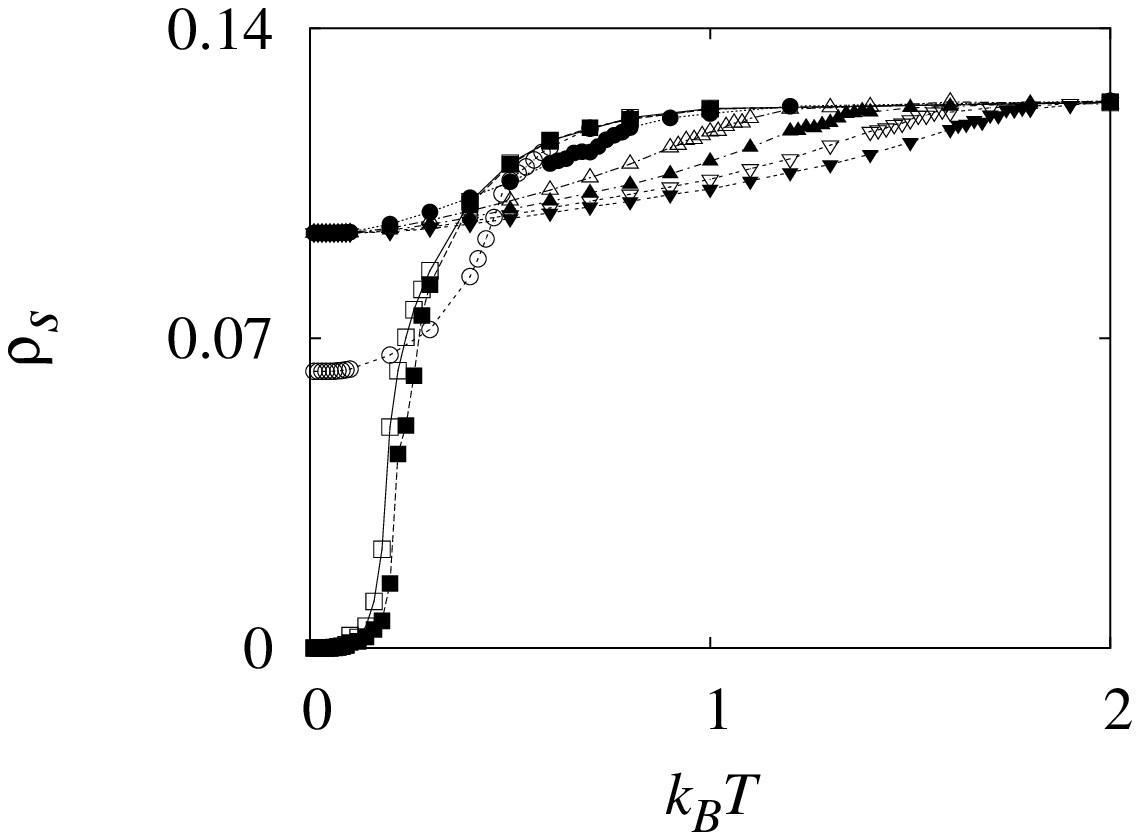}
    (c)
  \end{minipage}
  \begin{minipage}{8cm}
    \centering
    \includegraphics[width=\textwidth]{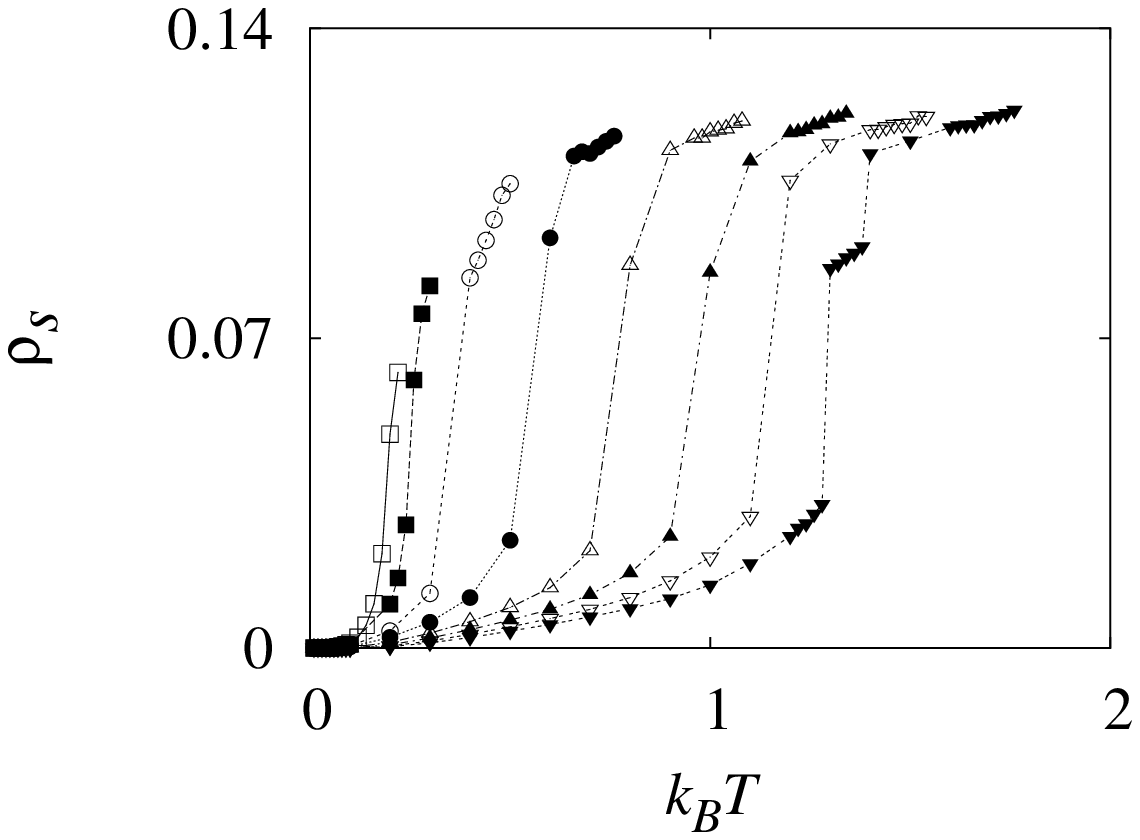}
    (d)
  \end{minipage}
  \caption{The order parameter $m$ [(a) and (b)] and the soliton density $\rho_s$ [(c) and (d)] versus the
    temperature $T$ in the $M=2$ system of misfit $\delta=0.4$ and size $L_z=32$.
    The left figures [(a) and (c)] and the right ones [(b) and (d)] show the data obtained in the cooling
    process and in the heating process, respectively.  Each symbol corresponds to a different value of the
    inter-chain coupling strength as listed in \figref{fig:ct}.}
  \label{fig:ccdw_d0.4}
\end{figure*}
\Figref{fig:ccdw_d0.4} compares the order parameter and the soliton density at $\delta=0.4$, measured
in two different processes: the cooling process and the heating one.
Unlike the case $\delta=0.2$, a hysteresis is evident at $\delta=0.4$ (and also in the undisplayed case of
$\delta=0.3$), as shown in \figref{fig:ccdw_d0.4}(a) and (b) for the order parameter and
in \figref{fig:ccdw_d0.4}(c) and (d) for the soliton density.
In the heating process staring from the commensurate ground state at zero temperature,
the commensurate CDWs survive some thermal fluctuations and experience the CI transition at a finite
transition temperature.  In contrast, as the temperature is lowered in the cooling process,
the order parameter keeps increasing and the soliton density saturates to a finite value,
unless the inter-chain coupling is sufficiently weak (i.e., for $U_\perp >0.02$).
This implies that the system still consists of the incommensurate CDWs at temperatures
where commensurate CDWs are supposed to be more stable.  For weak coupling ($U_\perp \le 0.02$),
on the other hand, no discrepancy between the two processes is observed.

It is of interest to compare the hysteresis observed here with the one reported in the specific heat
around the CI transition in a number of incommensurate systems.\cite{Hysteresis}
The latter is attributed to the effects of pinning of the incommensurate modulation due to
defects or impurities; our observation, in contrast, shows that the hysteresis can appear
even in the absence of defects, for which inter-chain interactions are responsible.
Namely, the inter-chain coupling operates in different ways depending on the process:
In the cooling process, correlations between CDW chains due to the inter-chain interactions
hinder each CDW chain from getting into the commensurate state.
On the contrary, in the heating process, the inter-chain correlations hold each chain close to
the zero-temperature ground state until thermal fluctuations become comparable to the inter-chain
interaction energy.  This argument is supported by the observation that the order parameter and
the soliton density increase abruptly in a narrow region of the temperature, as shown in
\figref{fig:ccdw_d0.4}(b) and (d), and such discrepancy becomes manifest as the inter-chain coupling
becomes stronger.  In addition, in the same narrow region the soliton number and the order parameter
fluctuate strongly while the energy fluctuations are relatively weak.  This suggests a kind of
configurational fluctuations over many metastable states.

For large values of the misfit ($\delta\gtrsim\delta_c$), neither commensurate CDW nor hysteresis
is observed and the soliton density does not vanish even at zero temperature in any process.
Only the transition separating the incommensurate CDW state from the disordered state is thus identified.

We have also performed similar simulations for the commensurability factor $M=3$, only to obtain
results qualitatively the same as those for $M=2$ presented above.  Quantitative differences observed
include that for $M=3$ the misfit affects the transition temperature more weakly
and the hysteresis takes place even at smaller misfit such as $\delta=0.2$.

%
%
\section{Phase Diagram\label{sec:pd}}

\begin{figure*}
  \centering
  \begin{minipage}{8cm}
    \centering
    \includegraphics[width=.75\textwidth]{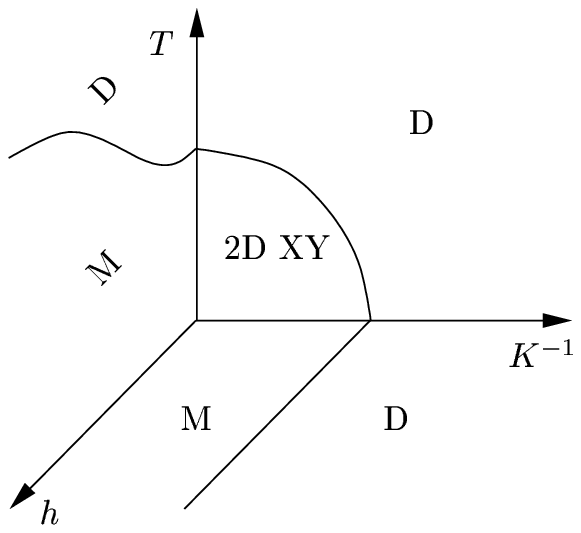}\\
    (a)
  \end{minipage}
  \begin{minipage}{8cm}
    \centering
    \includegraphics[width=.75\textwidth]{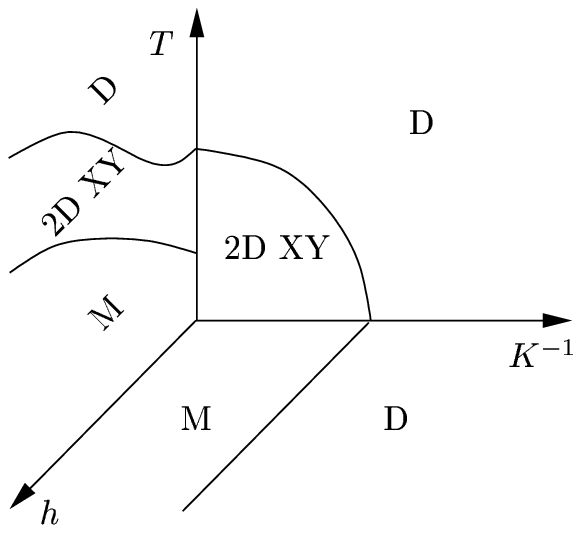}\\
    (b)
  \end{minipage}\\[0.5cm]
  \begin{minipage}{8cm}
    \centering
    \includegraphics[width=.75\textwidth]{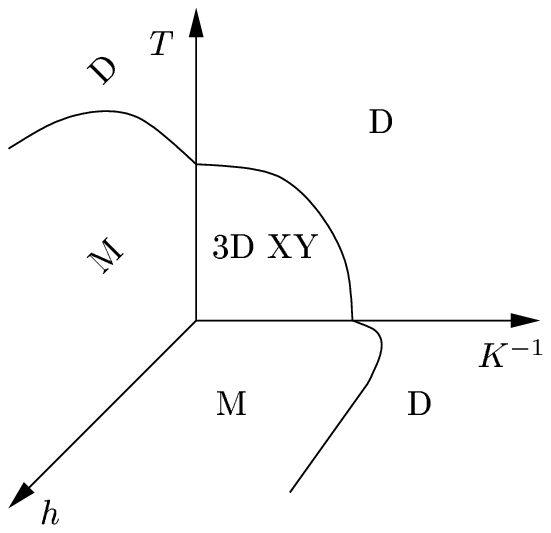}\\
    (c)
  \end{minipage}
  \begin{minipage}{8cm}
    \centering
    \includegraphics[width=.75\textwidth]{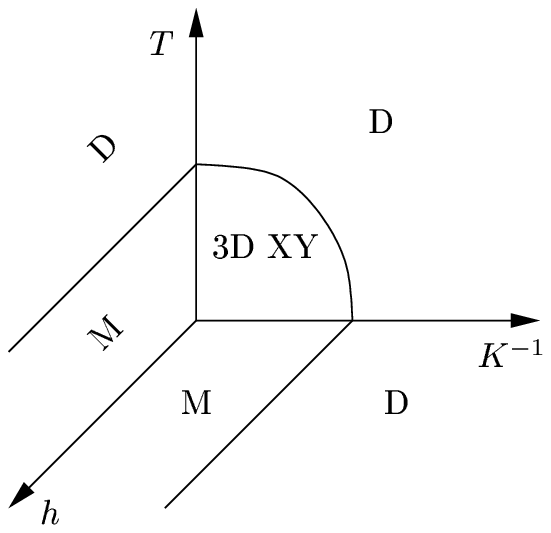}\\
    (d)
  \end{minipage}
  \caption{Schematic phase diagram for the coupled commensurate CDW system in the $(T, h, K^{-1})$
   space, depending on spatiotemporal fluctuations and commensurability:
   the homogeneous case (a) for $M=4$ and (b) for $M\ge5$; the general case (c) for $M=2$ or $3$ and
   (d) for $M\ge4$.  Plotted on each plane are boundaries between various phases, including the disordered
   phase (D), the $M$-state clock ordered phase (M), the algebraically ordered phase present in the 2D \XY
   model (2D XY), and the 3D \XY ordered phase (3D XY).}
  \label{fig:pd_xy}
\end{figure*}

Combining the results for the 3D and 4D \XY models under symmetry-breaking fields,
studied in \secsref{sec:3dxy} and \ref{sec:4dxy},
and those for the near-commensurate CDW model, studied in \secref{sec:cic},
we are ready to describe the phase transition in the coupled CDW system.
First, we focus on the system in the absence of any misfit $(\delta =0)$.
For this purpose, it is adequate to consider the 3D space consisting of the three parameters:
the temperature $T$, the strength of the symmetry-breaking field $h$,
and the amount of quantum or temporal fluctuations $K^{-1}$.
On each of the three planes ($T=0$, $K^{-1}=0$, and $h=0$) in the 3D space,
appropriate boundaries separating various phases can be plotted through the use of the known results,
as shown in \figref{fig:pd_xy}.  The phase at a given point in the 3D space can then be speculated
by means of finite-size scaling in the region near $T=0$ or the semi-classical methods
in the $K^{-1}\approx0$ region.

We first consider the homogeneous case, where spatial fluctuations along each chain are negligible,
and show the phase diagram in \figref{fig:pd_xy} for (a) $M=4$ and (b) $M\ge5$.
At zero temperature, the system is mapped onto the 3D \XY model as discussed in \secref{sec:ea},
and the corresponding phase diagram, obtained in \secref{sec:3dxy}, is sketched on the $h$-$K^{-1}$ plane.
On the other hand, in the absence of temporal fluctuations ($K^{-1}=0$), the system maps onto
the classical 2D \XY model under symmetry-breaking fields, for which the phase diagram in
Ref.~\onlinecite{Jose77} is drawn on the $h$-$T$ plane.  Finally, on the $T$-$K^{-1}$ plane with $h=0$,
the system is described by the Hamiltonian in \eqnref{eq:2dxy} and expected to display the
Berezinskii-Kosterlitz-Thouless transition renormalized by quantum fluctuations.\cite{JJ,Comment}
Note that the cases $M=2$ and $3$ are not shown here.  In this case presumably Ising/Potts critical lines
exist on the $h$-$T$ plane; however, it is not known how these lines connect up to the phase boundary
for $h=0$.

In the general case with spatial fluctuations present, we obtain the phase diagram shown in
\figref{fig:pd_xy}(c) and (d) for $M\le3$ and $M\ge4$, respectively.
Owing to the additional dimension along the chain direction, the system maps onto the 3D \XY model
in the classical limit, i.e., on the $h$-$T$ plane, whereas 4D \XY model is obtained at zero temperature.
Accordingly, on the $h$-$K^{-1}$ plane, the phase diagram of the 4D \XY model obtained in
\secref{sec:4dxy} is drawn.
It is observed that the 3D \XY ordered phase does not survive the field and is replaced by the $M$-state clock
ordered phase for $h\ne0$.  For $M\ge4$, however, the commensurability energy affects neither
the transition temperature nor the nature of the transition.

\begin{figure}
  \centering
  \includegraphics[width=8cm]{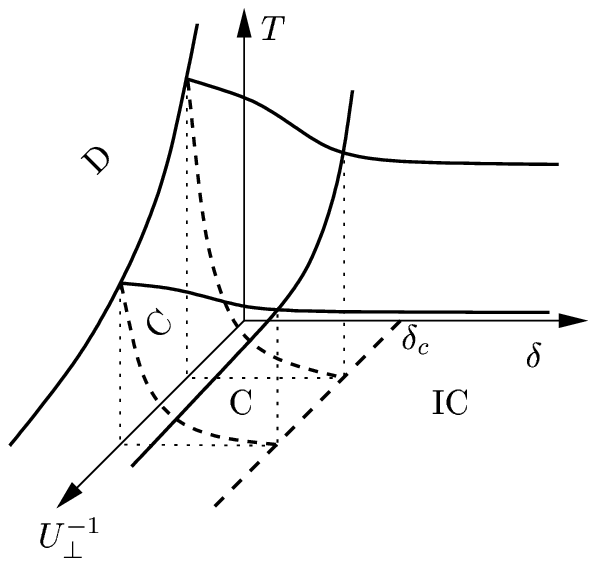}
  \caption{Schematic phase diagram for the coupled CDW system in the $(T, U_\perp^{-1}, \delta)$ space,
   with the effects of the misfit taken into account.  The surfaces depicted by thick solid lines
   and dashed lines separate the incommensurate CDW phase (IC) from the disordered phase (D) and
   from the commensurate CDW phase (C), respectively.}
  \label{fig:pd_ccdw}
\end{figure}

We next take into consideration the effects of misfit and show in \figref{fig:pd_ccdw}
the schematic phase diagram in the 3D space consisting of $T$, $U_\perp^{-1}$, and $\delta$,
with $U_\parallel$ and $V_0$ fixed.  At $\delta=0$, the system belongs to the same
university class as the 3D \XY model under the symmetry-breaking field,
for which the phase diagram is drawn on the $T-U_\perp^{-1}$ plane.
On the other hand, at zero temperature, each CDW chain undergoes the CI transition with the
critical misfit $\delta_c$ which is independent of the inter-chain coupling strength $U_\perp$.
The order-disorder transition temperature $T_D$ and the CI transition temperature $T_{IC}$
obtained in \secref{sec:cic} produce surfaces depicted by (thick) solid and dashed lines,
respectively, in the $(T,U_\perp^{-1},\delta)$ space.
The phase diagram shows that for $\delta<\delta_c$ the system undergoes double transitions
as the temperature is lowered: first, the order-disorder transition into the incommensurate
CDW phase and then the CI transition into the commensurate CDW phase.
As revealed in \secref{sec:cic}, a hysteresis takes place around the CI transition point,
which manifest itself more clearly as the inter-chain coupling becomes stronger.

%
%
\section{Conclusion\label{sec:c}}

To study phase transitions in coupled CDW systems, we have mapped the systems at zero temperature onto
three- or four dimensional \XY models, depending on the spatiotemporal fluctuations,
under symmetry-breaking fields which arise from the commensurability energy.
Such techniques as $\epsilon$-expansion, mean-field theory, and the Monte Carlo method
have then been applied to the obtained \XY models.  Revealed is a single second-order transition
between the $M$-state clock order and disorder in both the three-dimensional and four-dimensional
systems, except for the case $M=3$ in the four-dimensional system where the transition is of the first order.
In particular, the commensurability with $M\ge4$ has been observed not to change the properties of
the transition including the critical temperature and exponents.
Combining these zero-temperature ($T=0$) results with the existing results in the absence of quantum
(temporal) fluctuations ($K^{-1}=0$) or of the symmetry-breaking field ($h=0$), we have
constructed boundaries separating various phases on the three planes ($T=0$, $K^{-1}=0$, and $h=0$)
in the three-dimensional $(T, K^{-1}, h)$ space.  The boundaries near $T=0$ and near $K^{-1} =0$
can then be speculated through the use of finite-size scaling and semi-classical methods, respectively,
thus giving a schematic phase diagram in the $(T, K^{-1}, h)$ space.

We have also found via Monte Carlo simulations that the system with nonzero misfit undergoes
a commensurate-incommensurate transition.  The inter-chain interactions give rise to the correlations
between neighboring CDWs in such a way that either the commensurate state or the incommensurate
state is favored depending on the initial configuration: In the cooling process the CDWs remain
incommensurate down to almost zero temperature while in the heating process a substantial
amount of the commensurate CDWs survive thermal fluctuations.

At strong thermal or quantum fluctuations, i.e., at high $T$ or small $K$, the system is in the
disordered phase.  No CDW is formed and the system is expected to be metallic.  On the contrary,
weak fluctuations (low $T$ and large $K$) favor the $M$-state clock ordered phase,
in which commensurate CDWs are developed as long as $\delta<\delta_c$.
Accordingly, the interaction between the periodicity of the CDW and the underlying lattice periodicity
drives the collective excitation to develop a gap, and the system becomes insulating.
At moderate fluctuations or for large misfit ($\delta>\delta_c$), on the other hand,
incommensurate CDWs emerge.  In this case the system may remain conducting through
collective Fr\"ohlich conduction, i.e., via sliding of the CDWs.
Usually, the conductivity via such collective modes is lower than that via uncondensated electrons
in the disordered phase (without CDW).  In particular the CDW may be pinned in the presence of impurities,
sharply decreasing the conductivity.  As the temperature is lowered, therefore, the system for weak
quantum or thermal fluctuations becomes insulating via three possible routes, depending on
the misfit and the inter-chain interaction:
First, the commensurate CDW phase emerges directly from the high-temperature disordered phase;
second, only the incommensurate CDW phase appears, reducing the conductivity;
third, the incommensurate CDW phase appearing first is followed by the commensurate phase emerging
via the commensurate-incommensurate transition.

Note that beginning with the Hamiltonian in \eqnref{eq:chainH}, we have taken into account
only phase fluctuations and disregarded amplitude fluctuations.  The latter are in general irrelevant
in the RG sense, expected not to affect nature of the phase transition.  On the other hand,
there still lacks conclusive understanding of the dynamic properties in various phases.
It is thus desirable to consider the responses to external electromagnetic fields, and
for example, compute the conductivity, which can be obtained from the current or the average
momentum in the presence of appropriate misfit.
Detailed investigation of such dynamic responses is left for further study.

\section*{Acknowledgments}

We thank G.S. Jeon for helpful discussions and acknowledge the partial support from the Korea
Science and Engineering Foundation through the SKOREA Program and from the Ministry of Education
of Korea through the BK21 Program.

%
%


\begin{thebibliography}{99}
\bibitem{Comment}
  Here we disregard the possible existence of re-entrance, which has been a source of
  controversy.  See Ref.~\onlinecite{JJ}.
\end{thebibliography}
\end{document}